\documentclass[showpacs,aps,prl,floatfix,superscriptaddress]{revtex4}
\usepackage{graphicx}
\usepackage{citesort}
\usepackage{epic,eepic}
\usepackage{graphics}
\usepackage{epsfig}
\usepackage{subfigure}
\newcommand{\G}{{\cal G}}
\newcommand{\x}{{\bm x}}

\newcommand{\be}{\begin{equation}}
\newcommand{\ee}{\end{equation}}
\usepackage{amsmath}
\usepackage{amsfonts}
\usepackage{amssymb}
\usepackage{bm}
\usepackage{color}
\usepackage{graphicx}
\begin{document}

\title{Generalized Lattice Boltzmann Method with multi-range pseudo-potential}

\author{M. Sbragaglia} \affiliation{Department of Applied Physics, University of Twente, P.O. Box 217, 7500 AE Enschede, The Netherlands} 
\author{R. Benzi} \affiliation{Dipartimento di Fisica and INFN,
  Universit\`a di Roma ``Tor Vergata'', Via della Ricerca Scientifica
  1, 00133 Roma, Italy.}
\author{L. Biferale} \affiliation{Dipartimento di Fisica and INFN,
  Universit\`a di Roma ``Tor Vergata'', Via della Ricerca Scientifica
  1, 00133 Roma, Italy.}
\author{S. Succi} \affiliation{Istituto per le Applicazioni del
  Calcolo CNR, Viale del Policlinico 137, 00161 Roma, Italy.}
\author{K. Sugiyama} \affiliation{Department of Applied Physics, University of Twente, P.O. Box 217, 7500 AE Enschede, The Netherlands} 
\author{F.\ Toschi} \affiliation{Istituto per le Applicazioni del
  Calcolo CNR, Viale del Policlinico 137, 00161 Roma, Italy.}
\date{\today}

\author{} 
\date{\today}

\begin{abstract}
The physical behaviour of a class of mesoscopic models for multiphase flows 
is analyzed in details near interfaces.  In particular, an extended pseudo-potential method is developed, which 
permits to tune the equation of state and surface tension independently of each other.
The spurious velocity contributions of this extended model 
are shown to vanish in the limit of high grid refinement and/or high order isotropy. 
Higher order schemes to implement self-consistent forcings are rigorously computed for $2d$ and $3d$ models.
The extended scenario developed in this work clarifies the theoretical foundations
of the Shan-Chen methodology for the lattice Boltzmann method and enhances its applicability and flexibility
to the simulation of multiphase flows to density ratios up to  ${\cal O} (100)$.
\end{abstract}
\maketitle

\section{Introduction}

The Lattice Boltzmann method \cite{BSV,Chen,Gladrow} developed in the late
$80$'s as an efficient and powerful way to simulate nearly
incompressible hydrodynamics and its multiphase extensions
\cite{YEO,SC1,SC2} represent one of the most successful 
mesoscopic techniques for numerical simulation of complex flows.

Besides the mainstream application, namely complex macroscopic flows
far from equilibrium, recent work is also hinting at the possibility
that multiphase lattice Boltzmann methods may provide a new
methodological framework for the description of fluid-solid
interactions which play a crucial role for micro/nano-fluidic
applications \cite{Squires,Tabebook}. For example, the possibility to
model slip boundary conditions and wetting properties
\cite{Yeomans1,PREnostro,JFM,EPLnostro,Harting1,Harting2} has been
recently achieved within the framework of the lattice Boltzmann
equation. More detailed comparisons between the mesoscopic technique
and atomistic molecular dynamics simulations \cite{PRLnostro,Horbach}
have pointed out that lattice Boltzmann may become a method of choice for
physical problems where supramolecular details play a major role.
By supramolecular, we refer to situations which escape a purely
continuum treatment, and yet, still exhibit sufficient universality to
do away with a fully atomistic description. Arguably, a wide class of
multiphase flows out of equilibrium falls within this class.

All this looks promising, especially in view of recent experimental
activity aimed at shedding some light on the rich and still largely
unexplored territory of dynamical behaviour of liquids confined at (or
below) millimetric scales.  Impact of droplets on solid substrates,
droplets breakup, capillarity instabilities and bouncing transitions,
liquid fragmentation and water repellency on structured surfaces, are
just but a few examples in point
\cite{Yarin,Xu,Reyssat,Kim,Bartolo,Quere1,Quere2}.

Since the phenomenological description is not based on molecular
details but only on average properties (for example surface tension,
contact angle) mesoscale modelling and numerical simulations would be
extremely helpful to  access time and space scales of
direct experimental relevance.

This is confirmed by recent numerical simulations for static behaviour
\cite{PRLnostro,Yeomans1} and also by  some attempts to describe contact
line motions \cite{Briant1,Briant2,Kwok} and dynamical properties
induced by heterogeneous wetting \cite{Yeomans1,PREnostro}.\\

These recent developments unquestionably set a compelling  case for
revisiting and extending some basic theoretical aspects of multiphase
mesoscopic methods. In particular, the pseudo-potential approach
introduced a decade ago by Shan and Chen (SC) \cite{SC1,SC2} to deal
with non-ideal inhomogeneous fluids, represents one of the most
successful outgrowths of the Lattice Boltzmann theory.  It is worth
noticing that non-ideal fluid behaviour can also be encoded a-priori
by deriving lattice local equilibria directly from a free-energy
functional \cite{YEO}. This option leads to local equilibria with an
explicit dependence on the density gradients, which cannot be
readsorbed into a compact shift of the velocity, as it is the case for
the pseudo-potential method \cite{SC1,SC2}. The result is that the pseudo-potential method, 
albeit in-principle less rigorous, is very
flexible and robust for practical and numerical purposes.\\
Despite its undeniable success, this method has made the object of
extensive criticism, the major objections being that surface tension
is not tunable independently of the equation of state and that the
interface dynamics is
affected by spurious currents near (curved) interfaces.\\

\noindent In this paper, it is shown that both above limitations can
be lifted by moving to a mid-ranged pseudo-potential, i.e.  by
extending the spatial range of the pseudo-potential interaction.  More
specifically, it will be shown that (i) surface tension can be tuned
independently of the equation state, by formulating a two-parameter
version of the SC model with mid-range interactions, (ii) spurious
currents near curved interfaces become vanishingly small in the limit
of zero mesh-spacing and/or in the limit of an isotropic lattice. These
developments help to put pseudo-potential methods a-la Shan-Chen on a
solid theoretical basis.

\section{Mean field approach: Shan-Chen model and its generalizations}

In this section, we briefly recall the main features of the lattice
Boltzmann equation and the application to multiphase flow via the
introduction of a pseudo-potential. The main goal here is to
understand the corrections to the ideal-gas equations introduced by
the presence of  attractive pseudo-potential between Boltzmann
kinetic populations.

We start from the usual lattice Boltzmann equation with a single-time
relaxation \cite{LBGK,Gladrow,Saurobook}: 
\be\label{eq:LB}
f_{l}(\bm{x}+\bm{c}_{l}\Delta t,t+\Delta
t)-f_l(\bm{x},t)=-\frac{\Delta t}{\tau}\left(
  f_{l}(\bm{x},t)-f_{l}^{(eq)}(\rho,\rho {\bm u}) \right) +F_l 
\ee
where $f_l(\bm{x},t)$ is the kinetic probability density function
associated with a mesoscopic velocity $\bm{c}_{l}$, $\tau$ is a mean
collision time (with $\Delta t$ a time lapse), $f^{(eq)}_{l}(\rho,\rho
{\bm u})$ the equilibrium distribution, corresponding to the
Maxwellian distribution in the continuum limit and $F_l$ represents a
general forcing term whose role will be discussed later in the
framework of inter-molecular interactions.  From the kinetic
distributions we can define macroscopic density and momentum fields as
\cite{Gladrow,Saurobook}: 
\be \rho(\x)=\sum_{l} f_{l}(\x) \ee \be \rho
{\bm u}(\x)=\sum_{l}{\bm c}_{l}f_{l}(\x).  \ee 
For technical details
and numerical simulations we shall refer to the nine-speed,
two-dimensional $2DQ9$ model \cite{Gladrow}, often used due to its
numerical robustness \cite{Karlin}. The equilibrium distribution
in the lattice Boltzmann equations is obtained via a low Mach number
expansion of the continuum Maxwellian \cite{Gladrow,Saurobook}
\be\label{EQUILIBRIUM} f^{(eq)}_{l}={w}^{(eq)}_{l}\left[
  \rho+\frac{c^{i}_{l} \rho u_{i}}{{c}^{2}_{s}}+
  \frac{(c^{i}_{l}c^{j}_{l}-{c}^{2}_{s}\delta_{ij})}{2 {c}^{4}_{s}}
  \rho u_{i}u_{j} \right] \ee where ${{c}}^{2}_{s}=1/3$ and
$i=1,2=x,y$ runs over spatial dimensions.  The weights
${w}^{(eq)}_{l}$ are chosen such as to enforce isotropy up to fourth
order tensor in the lattice \cite{Gladrow,Saurobook}. From the
equilibrium distribution and the symmetry properties of ${\bm c}_l$, it
immediately follows \cite{Gladrow} the kinetic second order tensor of
the equilibrium distribution:
$$
\sum_{l}f^{(eq)}_{l}{c}^{i}_{l}{c}^{j}_{l}=\delta_{ij}({{c}}^{2}_{s}\rho)+ \rho u_{i}u_{j},
$$
where, in the first term of the rhs, we recognize the well-known
ideal-gas pressure tensor: \be\label{eq:ideal} P_{ij} =\delta_{ij}(
c_s^2).  \ee In order to study non-ideal effects we
need to supplement the previous description with an interparticle
forcing. This is done by choosing a suitable $F_l$ in (\ref{eq:LB}).
In the original SC model \cite{SC1,SC2}, the bulk interparticle
interaction is proportional to a free parameter (the ratio of
potential to thermal energy), ${\cal G}_{b}$, entering the equation
for the momentum balance: \be\label{forcing} F_{i}=-{\cal G}_{b}
c^{2}_{s}\sum_l w(|{\bm c}_{l}|^2) \psi(\x,t) \psi (\x+{\bm c}_l\Delta
t,t) {c}^{i}_l \ee being $w(|{\bm c}_{l}|^2)$ the static weights
($w(1)=1/3$, $w(2)=1/12$ for the standard case of 2DQ9 \cite{Gladrow})
and $\psi(\x,t)=\psi(\rho(\x,t)$ the (pseudo) potential function which
describes the fluid-fluid interactions triggered by inhomogeneities of
the density profile.  The only functional form of the pseudopotential
$\psi(\rho)$ strictly compatible with thermodynamic consistency is
$\psi(\rho)=\rho$ \cite{PREnostro,Hedoolen}.  For purposes which will
become apparent in the sequel, here we shall refer to the
pseudopotential used in the original SC work \cite{SC1}, namely \be
\label{PSI} \psi(\x)=(1-exp(-\rho({\bm x}))). 
\ee
Note that this reduces to the correct form $\psi \rightarrow
\rho$ in the limit $\rho \ll 1$, whereas at high density ($\rho \gg 1$), it shows
a saturation. This latter is crucial to prevent density collapse of the
high-density phases (note that the SC potential is purely attractive,
so that a mechanism stabilizing the high-density phase is mandatory to
prevent density collapse).  In principle, other functional forms may
be investigated, sometimes with impressive enhancement
of the density ratios supported by the model \cite{Yuan}.\\

In order to understand the corrections to the ideal-state equation
(\ref{eq:ideal})  induced by the pseudo-potential, we need to define a
consistent pressure tensor, $P_{ij}$, for the macroscopic variables:
\be
\label{eq:cons}
\partial_j P_{ij} \equiv -F_i +\partial_i (c_s^2 \rho).  \ee Upon
Taylor expanding the forcing term and assuming hereafter $\Delta t=1$,
we obtain \be
\label{taylor}
F_{i}=-{\cal G}_{b}\psi \partial_{i}\psi - \frac{{\cal
    G}_{b}}{2}\psi \partial_{i} \Delta \psi + {\cal O}(\partial^5) \ee
which is correctly translated into \be\label{TENSORE} P_{ij}=\left(
  c^{2}_{s}\rho+{\G}_{b}\frac{c_{s}^{2}}{2}\psi^{2}+{\G}_{b}\frac{c^{4}_{s}}{4}|{\bf
    \nabla}\psi|^{2} +{\G}_{b} \frac{c^{4}_{s}}{2}\psi \Delta \psi
\right) \delta_{ij}- \frac{1}{2}{{\G}_{b} {{c}}^{4}_{s}}\partial_{i}
\psi \partial_{j}\psi+ {\cal O} (\partial^4).  \ee Let us notice that there
exists a sort of {\it gauge-invariance} in the definition of the
pressure tensor, and (\ref{TENSORE}) is just one of these.  In fact,
while the term $\psi \partial_{i} \psi$ is uniquely written as the
gradient of $\psi^2/2$, the same is not true for the term
$\psi \partial_{i}\Delta \psi$.  There are infinitely many tensorial
structures that correspond to the same $\psi \partial_{i}\Delta \psi$.
On the other hand, from its very definition, it is clear that the
tensor $P_{ij}$ is defined modulo any divergence-free tensor.
However, it can be shown (see Appendix A) that all tensorial
structures consistent with the forcing yield the same macroscopic
surface tension and density profiles across the interface.  Dispensing
with consistency between the forcing term and the pressure gradient in
the continuum, several choices for the pressure tensor can be proposed
\cite{SC2}. Hereafter, we will stick to the requirement to have any of
the possible {\it gauge-invariant} definition of the forcing and use  the expression (\ref{TENSORE}) for all subsequent technical developments.\\

In order to calculate the density profile for a  flat interface in $2d$ whose dishomogeneities 
develop along a single coordinate, say $y$,  we follow the mathematical details discussed in \cite{SC2,PREnostro} 
and impose the mechanical equilibrium condition for the normal component $P_{yy}$ of the above pressure tensor 
\be
\label{nabla}
\partial_{y}P_{yy}=0 \ee with the boundary conditions that
$\rho(-\infty)=\rho_{g}$ and $\rho(+\infty)=\rho_{l}$ , being
$\rho_{g},\rho_{l}$ the densities of the two phases.  After some
lengthly algebra (see \cite{PREnostro} for all details) one can show
that the densities in the two phases are fixed by an integral
constraint \be
\label{eq:inte}
\int^{\rho_l}_{\rho_g} \left[P_{bulk}-c_s^2 \rho
  -\frac{c_s^2\G_b}{2}\psi(\rho)^2\right] \frac{\psi'}{\psi^2} d \rho=
0 \ee where $P_{bulk}$ defines the equilibrium bulk pressure in one of
the two phases, $P_{bulk} = c_s^2 \rho_l
+\frac{c_s^2\G_b}{2}\psi(\rho_l)^2 = c_s^2 \rho_g
+\frac{c_s^2\G_b}{2}\psi(\rho_g)^2 $.  Let us notice that expression
(\ref{eq:inte}) is different from equations (25)  of
\cite{SC2} (see also \cite{Hedoolen}) because the latter is derived
without imposing consistency with the forcing (\ref{eq:cons}).
Anyhow, for all practical purposes,the difference in the density
ratios between the two versions is fairly negligible (see figure 7 of
\cite{PREnostro}).\\

The surface tension also follows as the integral along a flat surface of the mismatch 
between the normal $P_{yy}$ and transversal $P_{xx}$ component of the pressure tensor\cite{Rowlinson}:
\be 
\sigma=\displaystyle\int_{-\infty}^{+\infty} (P_{yy}-P_{xx})
dy=-\frac{{\cal G}_{b} c_{s}^{4}}{2} \displaystyle\int_{-\infty}^{+\infty}|\partial_{y}
\psi|^{2} dy. 
\ee

\subsection{Pseudo-potential with mid-range interactions}

\begin{figure}
  \begin{center}
    \includegraphics[scale=.7]{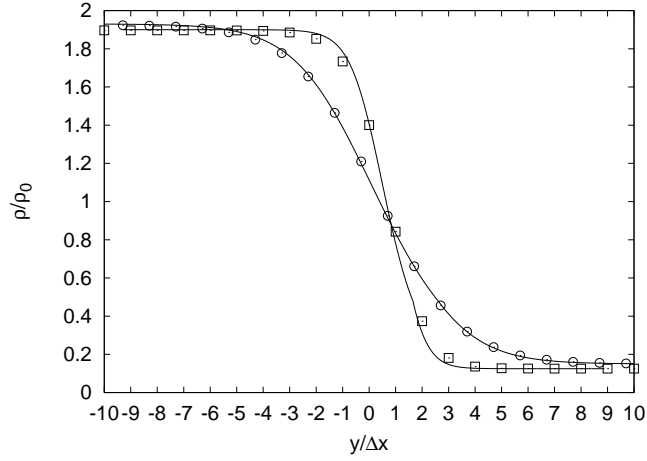}
    \caption{Smoothing of interface properties. We show the static
      profiles for a flat interface with an improved SC model
      (\ref{forcing2}) and (\ref{PSI}). The parameters are chosen to
      produce a macroscopic pressure tensor (\ref{TENSORENEW}) keeping
      fixed $A_{1}=-5.0$ (same density ratio) and at varying $A_{2}$:
      $A_{2}=-5.0$ ($\Box$) and $A_{2}=-30.0$ ($\circ$). The results
      of numerical simulations are compared with the analytical
      estimates (solid lines) obtained solving the mechanical
      stability equation (\ref{nabla}) applied to
      (\ref{TENSORENEW}). Notice the smoothing in the interface for a
      fixed density ratio due to a change in the surface tension.  In
      all numerical cases the lattice Boltzmann equation (\ref{eq:LB})
      has been integrated in time in a fully periodic domain $L_{x}
      \times L_{y}=100 \Delta x \times 100 \Delta x$ with a flat strip
      of liquid in an otherwise gaseous domain. The relaxation time is
      $\tau=0.7 \Delta t$.}
    \label{fig:1}
  \end{center}
\end{figure}

\begin{figure}
  \begin{center}
    \includegraphics[scale=.7]{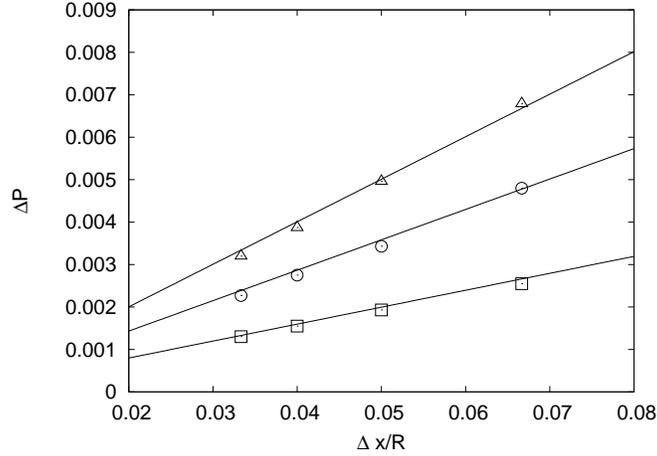}
    \caption{Laplace test for the same density ratio obtained from the
      numerical simulation of the improved SC model (\ref{forcing2})
      and (\ref{PSI}). The parameters are chosen to reproduce a
      macroscopic pressure tensor (\ref{TENSORENEW}) with $A_{1}=-5.0$
      and different $A_{2}$: $A_{2}=-5.0$ ($\Box$), $A_{2}=-15.0$
      ($\circ$), $A_{2}=-30.0$ ($\triangle$). For each case we plot the
      pressure drop $\Delta P$ as a function of the inverse radius of
      the static drop. Moreover we compare the results with the
      theoretical predictions (solid lines) given by the continuum analysis that
      leads to $\sigma(A_2= {-5})=0.0398$, $\sigma(A_2={-15})=0.0716$
      and $\sigma(A_2={-30})=0.100$ in lattice units. Numerical
      details are the same of figure (\ref{fig:1}) with the only
      difference that now we use a drop in the middle of the domain.}
    \label{fig:2}
  \end{center}
\end{figure}

It is immediately realized that, since in the SC model there is just
a single free-parameter, $\G_b$, it is impossible to tune density ratios
(i.e. equation of state) and surface tension (i.e interface width)
independently.  In order to discuss this problem, let us go back to the expression of
the forcing and consider possible generalizations thereof.
The  most immediate generalization of the standard SC model
reads as follows: 
\be\label{forcing2} 
F_{i}=-c_s^2 \sum_l w(|{\bm c}_{l}|^2) \psi(\x) [{\G}_{1} \psi
(\x+{\bm c}_l)+{\G}_{2} \psi (\x+2{\bm c}_l)] {c}^{i}_l, 
\ee
where interactions up to next-nearest neighbors are 
explicitly enabled. The corresponding equilibrium pressure-tensor takes
now the form:
 \be\label{TENSORENEW}
P_{ij}=\left( c^{2}_{s}\rho+A_{1}\frac{c_{s}^{2}}{2}\psi^{2}+A_{2}
  \frac{c^{4}_{s}}{4}|{\bf \nabla}\psi|^{2} +A_{2} \frac{c^{4}_{s}}{2}
  \psi \Delta \psi \right) \delta_{ij}- \frac{1}{2} A_{2}
c^{4}_{s} \partial_{i} \psi \partial_{j}\psi+ {\cal O} (\partial^4),  \ee
with $A_{1},A_{2}$ macroscopic constants related to
${\G}_{1},{\G}_{2}$ in (\ref{forcing2}):
 \be A_{2}={\G}_{1}+2 {\G}_{2}
\hspace{.2in} A_{1}={\G}_{1}+8 {\G}_{2}. 
\ee 
The surface tension now becomes
\be\label{sigmanew}
\sigma=\displaystyle\int_{-\infty}^{+\infty} (P_{yy}-P_{xx})
dy=-\frac{A_{2} c_{s}^{4}}{2} \displaystyle\int_{-\infty}^{+\infty}|\partial_{y}
\psi|^{2} dy,
\ee
with the profile obtained applying the mechanical 
stability equation (\ref{nabla}) to (\ref{TENSORENEW}).
 Let us first notice that 
the above two-parameters couplings can be viewed as the first two
terms of the expansion in terms of moments of the interacting potential: $A_p=
\sum_{n=1}^{n=\infty} n^p {\cal G}_p$, which is the lattice equivalent of the
continuum virial expansion: $A_p = \int dr r^p V(r)$, where $V(r)$ is a
 general atomistic interaction potential. 
In principle one could also enlarge the spectrum of mid-range interactions but, for our purposes, it is enough to consider a two
parameter coupling in (\ref{forcing2}). Infact, in the case of 
equation (\ref{TENSORENEW}) we have an expression depending on the two free-parameters,
$A_1,A_2$, and on the functional shape of $\psi$ as a function of
$\rho$. This opens up new degrees of freedom with respect to the standard SC formulation.
First, let us fix the pseudo-potential shape to:
\be
\psi(\rho) = \sqrt{\rho_0} (1-exp(-\rho/\rho_0))
\label{eq:fullrho}
\ee 
which reduces to the widely used choice \cite{SC2}, $\psi =
(1-exp(-\rho))$ for $\rho_0=1$. The importance of the free parameter
$\rho_0$ will become apparent later, when discussing the
grid-refinement of a given interface. For the moment we confine our
analysis to the standard case $\rho_0=1$ and we highlight the role of
the two parameters $A_1$, $A_2$ that, if used properly, allow to vary
the density separation $(\rho_l-\rho_g)/\rho_g = \Delta \rho/\rho_g$
between the two phases and the surface tension $\sigma$ independently
and in agreement with the continuum interpretation described through
(\ref{TENSORENEW}). In figures (\ref{fig:1}) and (\ref{fig:2}), we
show the equilibrium profiles obtained with a given $\Delta \rho
/\rho_g$ at changing the surface tension for both flat and curved
surfaces.  This is done using the two-parameters
forcing in such a way to reproduce the same $A_{1}$ but different
$A_{2}$ in (\ref{TENSORENEW}).  The numerical results for the case of
a flat interface are also in good agreement with the theoretical
predictions obtained from the mechanical stability equation,
$\partial_{y} P_{yy} =0$, applied to (\ref{TENSORENEW}).  To further check
the continuum interpretation given through (\ref{TENSORENEW}) we have
carried out Laplace tests for spherical droplets as shown in figure
(\ref{fig:2}).  From the Laplace law: \be \Delta P = P_{in}-P_{out} =
\sigma/R \ee being $P_{in}-P_{out}$ the difference between inner and
outer pressure in a spherical $2d$ droplet of radius $R$, we can
estimate the surface tension from a lin-lin plot of $\Delta P $ versus
$1/R$.  The numerically estimated surface tension agrees with the one
predicted by (\ref{TENSORENEW}), (\ref{nabla}) and
(\ref{sigmanew}). This is the first result presented in this paper. To
our knowledge such extension of the SC model leading to flexible
adjustment of the pressure tensor parameters, i.e. the surface tension
and the equation of state, has never been considered. This opens the
way to describe within a pseudo-potential approach more complex
physics where the surface tensions needs to be changed independently of
the equations of state as it is the case when, for example, surfactants are added changing the interface
properties \cite{Coveney} or when dynamical properties must 
be studied as a function of  $\sigma$ as for rising bubbles \cite{inamuro}.
\\
In the next subsection we use the extra-freedom given by the tunable
reference density $\rho_0$ in (\ref{eq:fullrho}) to change the
numerical resolution of the interface at fixed physics (i.e. fixed
surface tension and fixed density ratio). This is an important issue,
because of the inevitable numerical instabilities which limit the
density ratios obtained at a given spatial resolution.  Indeed, the
original SC model (\ref{forcing}) is known to be unable to describe
density jumps larger than $\Delta \rho/\rho_g \sim {\cal O}(10)$ per
grid point. This suggests the possibility of improving the flexibility
of the method by spreading the same density jump on a larger number of
grid nodes.

\subsection{Grid refinement and continuum description}

The introduction of $\rho_0$ in  (\ref{eq:fullrho}) 
allows us to refine the interface resolution for fixed density ratio
and surface tension. If we introduce the shorthand notation:
\be
\label{PSINEW}
{\psi}(\rho)=\sqrt{\rho_{0}} \tilde \psi(\tilde \rho); \hspace{.2in}
\tilde \psi (\tilde \rho ) = (1-exp(-\tilde \rho))
\ee
where $\tilde \rho = \rho/\rho_0$, the pressure tensor (\ref{TENSORENEW}) takes the following expression:
\be\label{eq:rescaled}
P_{ij}= \rho_0\left[
\left({{c}}^{2}_{s}\tilde \rho+A_{1} \frac{{{c}_{s}}^{2}}{2} {\tilde
    \psi}^{2}+A_{2}  \frac{{{c}}^{4}_{s}}{4}|{\bf \nabla}
  {\tilde \psi}|^{2} +A_{2}  \frac{{{c}}^{4}_{s}}{2} {\tilde
    \psi} \Delta {\tilde \psi} \right)\delta_{ij}-\frac{A_{2}
 {{c}}^{4}_{s}}{2}\partial_{i} {\tilde \psi} \partial_{j}
{\tilde \psi}\right]+{\cal O}(\partial^4). 
\ee
With reference to the case of a  flat interface with dishomogeneities only along the $y$ coordinate,  
by performing  the coordinate rescaling  $ y' = \lambda y $ in (\ref{eq:rescaled})  we obtain:
\be
\label{eq:rescaled2}
 P_{ij}=\rho_0\left[
\left({{c}}^{2}_{s}\tilde \rho +A_{1}\frac{{{c}_{s}}^{2}}{2} {\tilde
    \psi}^{2}+A_{2} \lambda^2  \frac{{{c}}^{4}_{s}}{4}|{\bf \nabla'}
  {\tilde \psi}|^{2} +A_{2} \frac{{{c}}^{4}_{s}}{2} \lambda^2
  {\tilde
    \psi} \Delta' {\tilde \psi} \right)\delta_{ij}-\frac{A_{2}
  {{c}}^{4}_{s}}{2} \lambda^2 \partial_{i}' {\tilde \psi} \partial_{j}'
{\tilde \psi} \right]+{\cal O}(\partial^4)
\ee
where $\nabla'$, $\Delta'$  and $\partial'$ means derivatives with
respect to the new variable, $y'$.
Let us notice that by choosing
\be
\label{eq:A2}
A_2 =  A_2'/\lambda^2 
\ee 
with $A_2'$ constant, the dependency on $\lambda$ disappears from
(\ref{eq:rescaled2}) and the only dependency on $\rho_0$ 
in the above expression comes from the overall prefactor. Therefore,
in the expression of the mechanical stability condition for a  flat interface 
 $\partial_y'P_{yy} =0$, as applied to (\ref{eq:rescaled2}), 
no dependence on $\rho_0$ and $\lambda$ is left. 
In this way, we are able to extract a universal
profile, $\tilde \psi (\tilde \rho)$ as a function of $y'$.
This leads to the conclusion that the density ratio $\Delta
\rho/\rho_g$ is independent of $\rho_0$.\\
 As for the surface tension,
equation (\ref{eq:rescaled}) in the old variables, yields:
\be
\label{eq:surface}
\sigma=-\frac{A_{2} \rho_0 }{2 {c}^{4}_{s}}\displaystyle\int
_{-\infty}^{+\infty}|\partial_{y} \tilde \psi|^2 dy
\ee 
which, in terms of $A_2'$ and $y'$,  becomes:
\be
\label{eq:surface2}
\sigma=-\frac{A_{2}' \rho_0 }{2 \lambda  {c}^{4}_{s}}\displaystyle\int
_{-\infty}^{+\infty}|\partial_{y'} \tilde \psi|^2 dy'.
\ee 
Since the profile and its integral in the {\it primed} variables 
is universal, from (\ref{eq:surface2}) we see that, by  choosing 
$$ \lambda = \rho_0 $$ 
the surface tension is also invariant 
under rescaling of the spatial coordinate.

\begin{figure}
  \begin{center}
    \includegraphics[scale=.7]{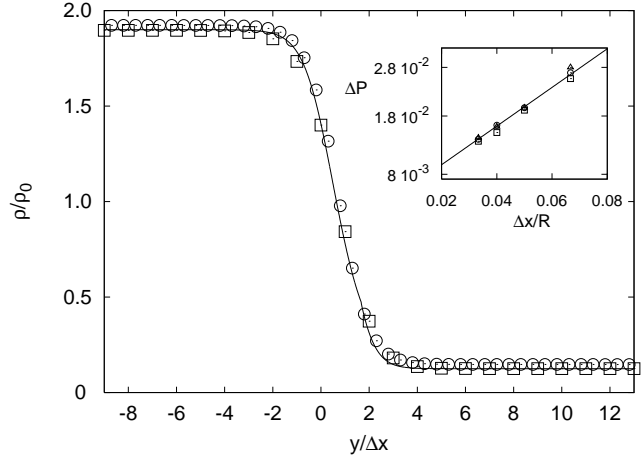}
    \caption{Smoothing of interface properties. We show the static
      profiles for a flat interface with an improved SC model
      (\ref{forcing2}) and (\ref{PSINEW}). The parameters are chosen
      such as to produce a macroscopic pressure tensor
      (\ref{eq:rescaled}) with $A_{1}=-5.0$ and
      $A_{2}=-5.0/\rho^2_{0}$ for different values of $\rho_{0}$:
      $\rho_{0}=1.0$ ($\Box$), $\rho_{0}=0.5$ ($\circ$) .  Results
      are also compared with the analytical estimate resulting from
      our continuum interpretation (solid line). The two profiles have
      been plotted by rescaling the lattice grid by a factor
      $1/\rho_0$. Notice the increased interface grid resolution
      obtained at fixed density ratio. Inset: the results of the
      Laplace tests made on spherical droplets with the same density
      ratio and varying grid resolution is also plotted. Again, we get
      the same surface tension with different interface resolutions,
      $\rho_{0}=1.0$ ($\Box$), $\rho_{0}=0.7$ ($\circ$),
      $\rho_{0}=0.5$ ($\triangle$). In all numerical cases the lattice
      Boltzmann equation has been integrated in time in a fully
      periodic domain $L_{x} \times L_{y}=100 \Delta x \times 100
      \Delta x$ with a flat strip of liquid in an otherwise gaseous
      domain. The relaxation time is $\tau=0.7 \Delta t$.}
    \label{fig:3}
  \end{center}
\end{figure}

It is therefore clear that $\rho_0$ in the functional form
(\ref{PSINEW}) can be used to fine-tune the thickness of the flat
interface at fixed values
of the physical parameters (density ratio and surface tension)  provided that we choose $A_{2}={A_2^{'}}/\rho^{2}_{0}$.\\
In figure (\ref{fig:3}), we show the equilibrium flat profiles for the
case (\ref{TENSORENEW}) and (\ref{PSINEW}), with
$A_1=-5.0$,$A_{2}=-5.0/\rho^2_{0}$ and different values of
$\rho_{0}$. As one can see the net effect is to change and magnify the
interface width with a good agreement with the analytical profiles
obtained from the continuum description given above.  We also carry out 
(see inset of figure (\ref{fig:3})) Laplace tests for the case
(\ref{TENSORENEW}) and (\ref{PSINEW}), with
$A_1=-5.0$,$A_{2}=-5.0/\rho^2_{0}$ and three different values of
$\rho_{0}$.  The macroscopic analysis predicts the same surface
tension and indeed this is precisely what the numerical simulations
show.  When moving from large to small $\rho_{0}$, a refinement of
the interface occurs.  Thus, fine-tuning of $\rho_0$ can be regarded
as a means of locally magnifying the interface region without changing
the macroscopic physics.

\section{Equilibrium description through Lattice Boltzmann Equations}

Up to now, we have mainly investigated the equilibrium properties of
interfaces resulting from the addition of a pseudo-potential in the 
classical Lattice Boltzmann formulation. A crucial point is
however, to analyze the dynamical stability of such results and to
understand the effects of the kinematic terms on the equilibrium
properties between the two phases. For weakly inhomogeneous fluids,
this is commonly achieved via the standard Chapman-Enskog expansion
\cite{Gladrow,Saurobook}  using the Knudsen number (molecular mean-free path over
smallest macroscopic scale, i.e. the width of the interface) as a
smallness parameter.  However, in the vicinity of a sharp-interface the Knudsen
number has no reasons to be small, being proportional to density
gradients, and the Chapman-Enskog procedure goes under
question. Recent work in this direction \cite{Buick,Guo,Ladd} has
carried out standard Chapman-Enskog analysis with additional forcing terms.
The proposed analysis leads to  a set of different macroscopic dynamic equations.
The correctness of the macroscopic limit is not analyzed
here. Infact, besides detailed {\it analytical} control on the behaviour of the
hydrodynamic fields close to the interface, one may  wonder whether 
numerical implementation of the lattice Boltzmann equation with a
pseudo-potential provides  realistic and stable results  over a wide range of 
density variations and surface tensions. 

Indeed, a disturbing phenomenon, known as {\it spurious currents}
\cite{Cristea,Wagner,Shan}, develops systematically in the vicinity of
interfaces: small circulating currents that are directly
proportional to the interface surface tension ({\it i.e.} density
ratio) spoil the physical results of numerical simulations and degrade
the numerical stability for high density ratios, 
thus casting serious doubts on the applicability of the method.\\

For flat interfaces, the situation is more under control.  In fact,
all spurious contributions reported near flat interfaces are due to an
ambiguity in the definition of the fluid momentum.  The correct way to
measure it,  is to take an averaged momentum between pre and 
post-collisional states \cite{Buick}.

This cures flat interfaces, but curved interfaces are still affected
by the problem and several attempts to justify and explain the origin
of this phenomenon have been proposed. In \cite{Cristea}, the author
proposed an {\it ad-hoc} extra counter-term to erase spurious
currents. Unfortunately, this analysis is limited to flat interfaces
and the prescription to erase the spurious currents is clearly
equivalent to averaging pre and post collisions momentum in the SC
model. In \cite{Wagner}, the author concluded that the origin of the
spurious currents is the incompatibility between the discretization of
the driving forces for the order parameters and momentum
equations. More recently, in \cite{Shan}, it has been shown that
spurious currents are due to insufficient isotropy of the discrete
forcing operator. In the latter paper, clear numerical evidence
is brought up, but no detailed analytical explanation is provided.\\
Here, besides supporting the numerical findings of \cite{Shan}, we
discuss in details the physical origin  of the spurious
currents. Then, following the symmetry analysis of lattice gas given
in \cite{Wolfram}, we derive improved isotropic schemes for $2d$ and
$3d$ models as well as further possible theoretical improvements.\\
The case of flat interface is pretty straightforward. In this case, let us denote  
again  with $y$ the direction of the non-homogeneity. 
We can imagine to have two homogeneous bulk phases
$\rho = \rho_g$ at $y=-\infty $ and $\rho = \rho_l$ at $y=+\infty$,
separated by an interface centered at $y=0$.  Then, the mass
conservation, $\partial_{t} \rho +\partial_{y} (\rho u_{y})=0$,   
in a stationary state ($\partial_{t} \rho=0$)  predicts $\rho  u_y = const$, independently of the
local density gradients, i.e. independently of the Chapman-Enskog
expansion \cite{Gladrow}.  Therefore, by imposing a zero net mass-flux at
infinity, one readily derives that $u_y=0$ everywhere.\\

\begin{figure}
  \begin{center}
    \includegraphics[scale=.7]{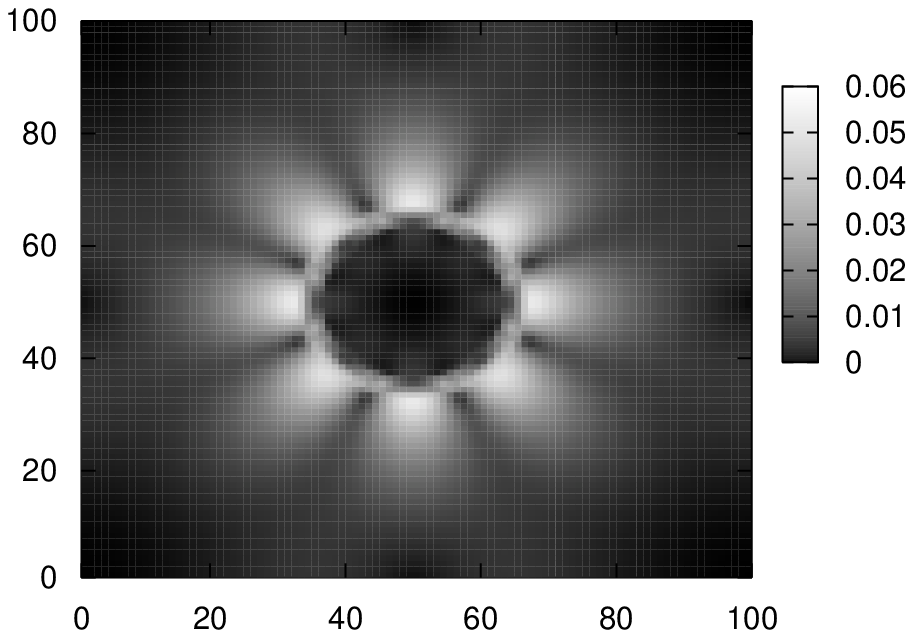}
    \includegraphics[scale=.7]{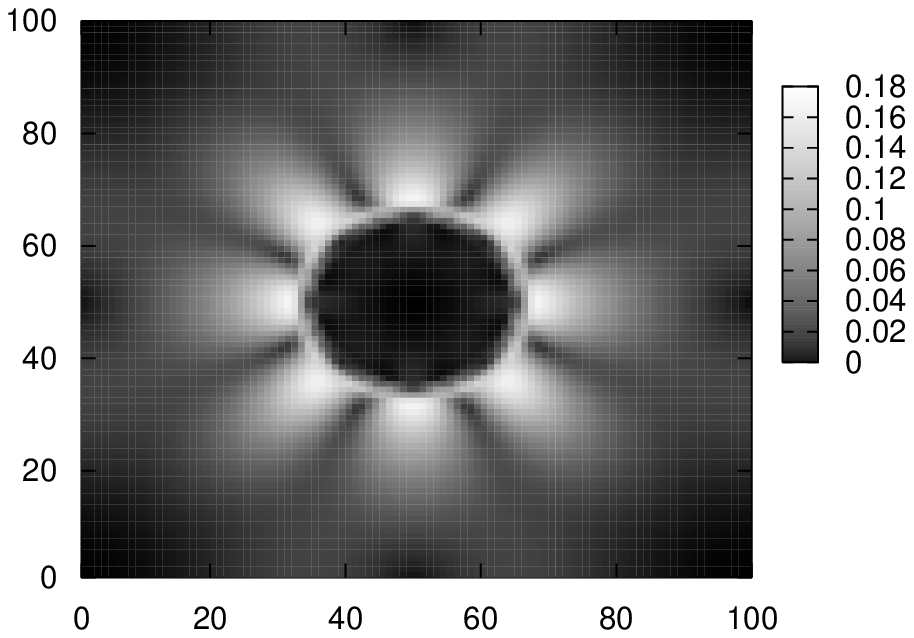}
    \caption{Spurious contributions around a static $2d$ drop for
      different density ratios with the standard SC model
      (\ref{forcing}) and (\ref{PSI}). Left: we show the local Mach
      number defined as $\sqrt{u^{2}_{x}+u^{2}_{y}}/c_{s}$ with
      $c_{s}$ the lattice sound speed. A drop of radius $15 \Delta x$
      and density ratio $\Delta \rho/\rho_{g} \sim 35$, obtained with
      ${\cal G}_{b}=-6.0$ in (\ref{forcing}), is considered. Right:
      the same as in the case of left figure with a higher density
      ratio $\Delta \rho/\rho_{g} \sim 60$ obtained with ${\cal
        G}_{b}=-7.0$ in (\ref{forcing}). Note in both plots the
      angular dependency due to lack of perfect radial symmetry in the
      forcing terms.  In all numerical cases the lattice Boltzmann
      equation (\ref{eq:LB}) has been integrated in time in a fully
      periodic $2d$ domain $L_{x} \times L_{y}=100 \Delta x \times 100
      \Delta x$ with the drop put in the middle of the system. The
      relaxation time is $\tau=1.0 \Delta t$.}
    \label{fig:4}
  \end{center}
\end{figure}

Let us now analyze the case of a circular drop in 2 dimensions.  The
new feature is that fluctuations tangential to the surface may also
appear and their connection with the forcing term  plays a key
role.  Infact, if the forcing is perfectly isotropic: \be
\label{FFF} F_i({\bm x})=\frac{x_i}{r}\tilde{F}(r)={\bm e}_{r}\tilde{F}(r), 
\ee 
where $\tilde{F}$ is a scalar function and ${\bm e}_r$ is the unit radial 
vector, one would argue that the velocity
field reflects the same symmetry, i.e. no spontaneous breaking of
rotational invariance should arise. For a stationary state, the
mass conservation implies: $r \rho  u_r = const.$, being $u_{r}$ 
the radial component of the velocity field. The only physical 
acceptable solution is $u_r=0$ everywhere. We conclude that if
the isotropy of the problem is perfectly carried over by the
discretization scheme, no spurious currents would develop even for a
curved interface. As a consequence, the numerically observed currents
must be stem from a lack of isotropy at some level with the  main contribution 
to anisotropy near the interface due to the pseudo-potential. Indeed,
one notices that
according to the set of grid points and weights entering in the
simplest expression of the forcing (\ref{forcing}), one has a loss of isotropy at a given order in
the Taylor expansion. For example, for the simplest case of 2DQ9
one obtains (see Appendix B for details):
\be 
\sum_l w_{l}(|{\bm c}_l|^2)\psi({\bm x}+{\bm c}_l){\bm c}_l =
{\bm \nabla} \left( 1+\frac{1}{6}\nabla^2 +\frac{1}{72}\nabla^2\nabla^2
\right)\psi +{\bm e}_x\frac{\partial_x^5\psi}{180} +{\bm
  e}_y\frac{\partial_y^5\psi}{180} + {\cal O}(\partial^7),
\label{eq:wpsic02now}
\ee where ${\bm e}_x=(1,0)$ and ${\bm e}_y=(0,1)$ are unit vectors in
Cartesian coordinates. If we have axial symmetry of the density
distribution we must have: $\psi=\psi(r)$. Because of $ {\bm \nabla}=
{\bm e}_r\partial r$ and $\nabla^2=\partial_r^2+r^{-1}\partial_r$ for
an axially symmetric function, the first term in the rhs of
(\ref{eq:wpsic02now}) is isotropic.  On the other hand, the $2$nd and
$3$rd terms, that arise only at the fifth order, are manifestly
anisotropic.  We should also notice that in the previous section, we
limited our analytical analysis to the 4th order expansion, and all
the numerical comparison where made by checking that spurious currents
arising from higher orders were negligible, since we chose a
stationary regime with small local gradients in the density field.
Nevertheless, on the route to higher density ratios, i.e. for cases
with high local density gradients, one necessarily meets with the
problem of anisotropic contributions.  In figure (\ref{fig:4}) we show
the structure of the spurious currents for two cases. As one can see,
the currents exhibit typical anisotropies with a quadrupolar
modulation, the result of anisotropies induced by higher order
derivatives in the pseudo-potential expansion (\ref{eq:wpsic02now})
and they are enhanced systematically when the density separation
between the two phases is increased.
  
\begin{figure}
  \begin{center}
    \includegraphics[scale=.7]{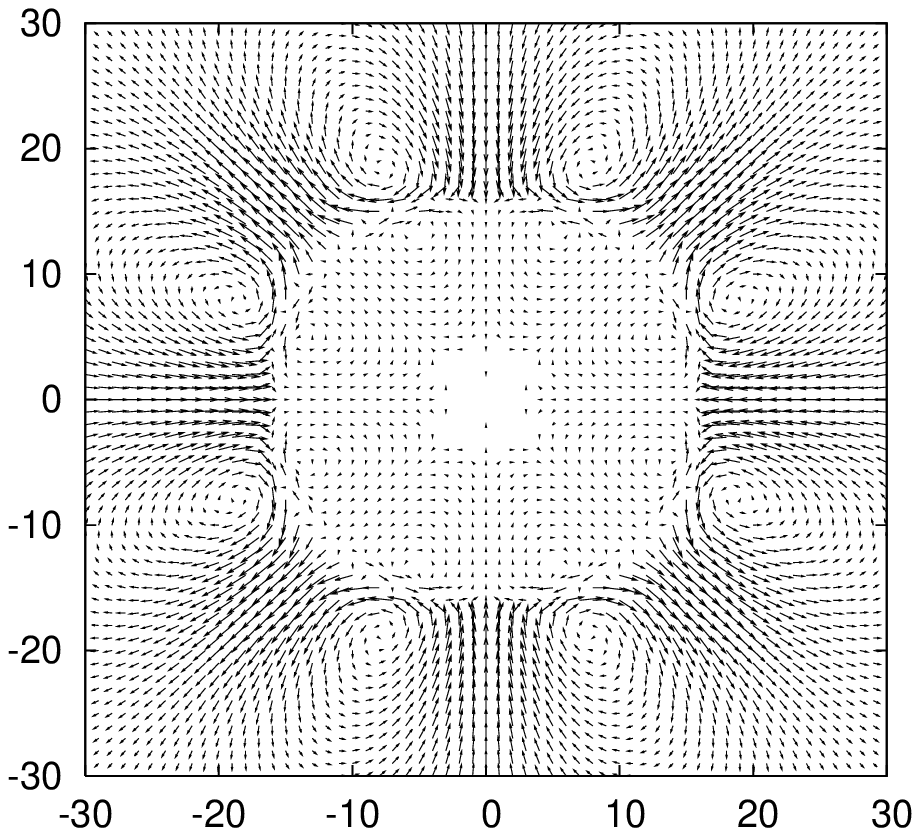}
    \includegraphics[scale=0.7]{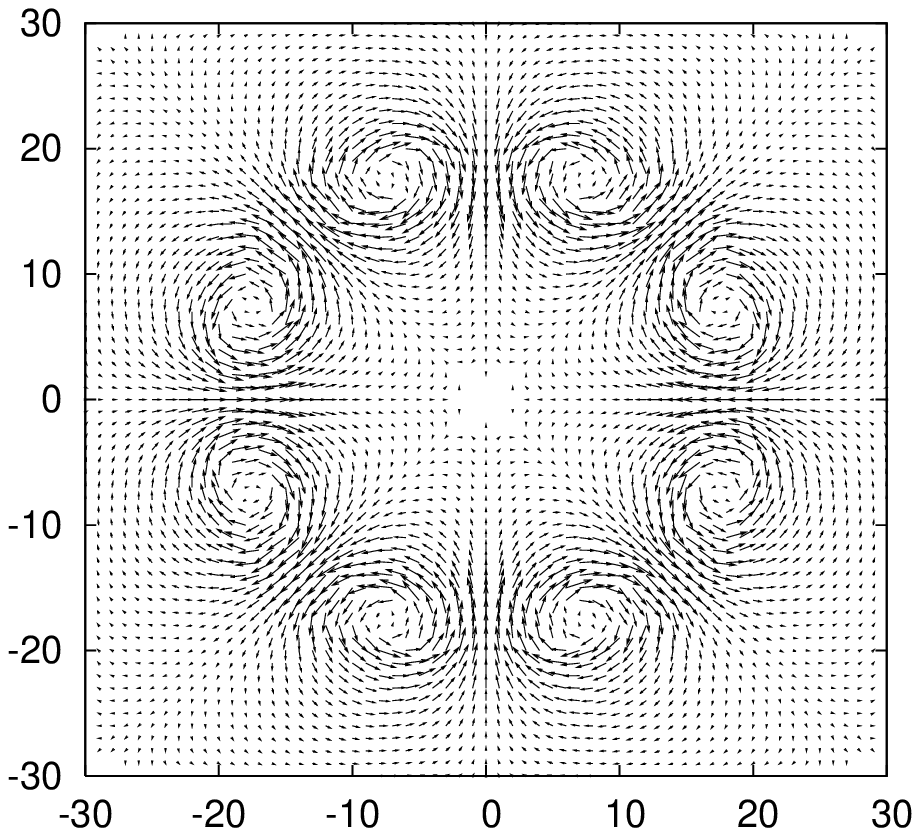}
    \caption{Spurious contributions in Lattice Boltzmann and their
      continuum interpretation. Left: we report the structure of the
      velocity field around a static drop of radius $15 \Delta x$ and
      density ratio $\Delta \rho/\rho_{g} \sim 50$. The data are the
      same of right panel of figure (\ref{fig:4}). Right: for a
      qualitative comparison we have solved the Laplace equation
      together with the continuity equations with a matching condition
      in a radial ring of a given width. The velocity field is
      obtained in an iterative way by first solving the Poisson
      equation $\nabla^2 \phi=\nabla \cdot {\bm u}$ and then renewing
      the velocity filed as ${\bm u}\rightarrow {\bm u}-\nabla
      \phi$. For both ${\bm u}$ and $\phi$ periodic boundary
      conditions are imposed in the horizontal and vertical
      directions. On the ring of range $17.5 \Delta x < r < 20.5
      \Delta x$ the matching condition: $u_{r}=\mbox{cos} 4 \theta,
      u_{\theta}=0 $ is imposed being $u_{r}$ and $u_{\theta}$ the
      radial and azimuthal components of the velocity field
      respectively.}
    \label{fig:5}
  \end{center}
\end{figure}

To further support the previous statement, we have solved the Laplace
equation, $\Delta {\bm u} =0 $ with anisotropic boundary conditions
on a ring, $u_r = cos(4\theta)$,$u_{\theta}=0$ for $ r_1<r<r_2$ (see caption of figure (\ref{fig:5}) 
for details).  
The result is a non-zero profile in the bulk regions.  This is also compared
qualitatively with the spurious currents picture from a stationary state of a 
numerical simulation and a good qualitative agreement is observed (see figure
(\ref{fig:5})).  From these pictures, we see that spurious
currents, once generated on the interface, spread through the bulk
regions, thereby corrupting the physical content of numerical
simulations.

Having assessed that spurious currents are triggered 
by high-order angular harmonics due to lack
of sufficient isotropy, it is natural to seek new models with a higher degree of isotropy.
There are at least two parallel ways to remove this problem.  Either
one improves the support  of the underlying lattice structure
coupled by the pseudo-potential terms, so as to push anisotropy to higher and
higher Taylor orders, or one can keep a given degree
of isotropy of the forcing term  and  improved grid resolution, so that
curved surfaces become more and more refined, hence subject
to smaller local density gradients. 

\subsection{Isotropy at a fixed discretization}

The former kind of technical improvement has already been proposed by
\cite{Shan}. Here we support these previous findings, and we extend them
systematically to higher orders in full details for both $2d$ and $3d$ cases (see Appendixes C and D). 
Following \cite{Wolfram}, the key idea consists of enlarging only the set of spatial grid points
coupled by the pseudo-potential $\psi$ and choosing the appropriate
weights to enforce isotropy up to the desired order. 
For any practical purpose, one writes
\be
\label{eq:iso}
F_i=-{\cal G}_{b} c^{2}_{s}\psi({\bm x})\sum_l  w(|{\bm c}_l|^2)\psi({\bm x}+{\bm c}_l)c_l^i
\ee
where ${\bm c}_l$ runs over a given set of grid points, changing
according to the required order of isotropy (see figure (\ref{fig:6})) .
In fact, by applying the Taylor expansion (all details in appendix C) to 
(\ref{eq:iso}), one obtains: 
\begin{equation}
\begin{split}
F_{i}=
-\G_{b} c_s^2\psi({\bm x}) 
\Biggl[E^{(2)}_{ij}\partial_{j}\psi
+\frac{1}{3!}E^{(4)}_{ijkl}\partial_{jkl}\psi
+\frac{1}{5!}E^{(6)}_{ijklpq}\partial_{jklpq}\psi
+\frac{1}{7!}E^{(8)}_{ijklpqst}\partial_{jklpqst}\psi
+
.. \Biggr],
\label{eq:f_tayl01}
\end{split}
\end{equation}
with
 \be E^{(m)}=E^{(m)}_{i_{1}i_{2}...i_{m}}=\sum_{l}w(|{\bm
  c}_{l}|^2){\bm c}^{i_{1}}_{l}{\bm c}^{i_2}_{l}...{\bm c}^{i_m}_{l}
\ee and (obviously) zero odd tensors: \be E^{(2n+1)}=0.  \ee The
weights can be chosen in such a way to recover isotropy to the desired
order (see appendixes C and D). Clearly, more velocities are needed in
the implementation of the forcing terms (see figure (\ref{fig:6})).
Numerical results (see figure (\ref{fig:7})) do confirm a decay of the
magnitude of the spurious contributions as the order of isotropy is
raised.  Although the practical implementation of higher-order scheme
might not be as straightforward as the standard SC case, it is
nonetheless reassuring to know that a well-defined procedure to tame
spurious currents is available.

\subsection{Refinement at a fixed degree of isotropy}

Since non-isotropic terms in the standard SC
 forcing scale with fifth-order derivatives, it
is plausible to expect that these terms can be attenuated also by a refinement of the
interface resolution, i.e. by the rescaling procedure previously illustrated.
In fact, in the standard formulation (\ref{eq:wpsic02now})  
the spurious contributions are induced by the terms ${\bm e}_x\frac{\partial_x^5\psi}{180}$  
and ${\bm e}_y\frac{\partial_y^5\psi}{180}$ 
that should fade away by a progressive refinement of the grid.\\

In figure (\ref{fig:8}) we show how refining the grid for a fixed
surface tension does indeed decrease the amplitude of spurious
velocities.  Using (\ref{TENSORENEW}) and (\ref{PSINEW}), with the
scaling $A_{2} \sim 1/\rho^2_{0}$, the macroscopic system stays the
same: same surface tension and same density ratio. The only difference
is a net reduction of the spurious velocity.  Let us notice that the
improvement due to grid-refinement within the extended
pseudo-potential (\ref{forcing2})  with pressure tensor
(\ref{TENSORENEW}) seems more effective than the one induced by
high-order isotropic forcing in the original SC model. Indeed,
comparing figure (\ref{fig:7}) and (\ref{fig:8}) one notices that in
the latter an almost complete depletion of spurious currents is
obtained already with a simple factor $2$ in the rescaled coordinate. On the other hand, to reach similar level of accuracy in the original SC model one needs to improve the isotropy of the forcing up to order $10$ or even more. \\
The fact that the smoothing of the density profile permits to reduce
considerably spurious contributions allows to achieve quite large
density ratios, up to the order of $\Delta \rho/\rho_{g} \sim 100$, as
shown in figure (\ref{fig:9}), where we plot the maximal spurious
velocity $|U|_{max}$ normalized to the sound speed as a function of
the density ratio.\\
Of course, one may also imagine to combine the two proposals, using
the extended formulation (\ref{forcing2}) with higher degrees of
isotropy. Whether the numerical effort is worthwhile has to be decided
on a case-by-case basis. \\

\section{CONCLUSIONS}

The SC model is one of the most successful spinoffs of lattice
Boltzmann theory.  It has nonetheless made the object of extensive
criticism over the last decade \cite{Joseph}.  Part of this criticism
is simply misplaced, some other is not.  In particular, lack of
thermodynamic consistency, surface tension tied-down to the equation of
state, and spurious currents near sharp interfaces, have spurred
doubts on the applicability of the SC method to the simulation of
realistic multi-phase flows.  In this paper we have elucidated the
physical reasons behind the above weaknesses, and also suggested
practical ways around them in the large-scale limit.

First, we have shown that by enlarging the number of coupling terms in the
pseudo-potential expression, one can push the method at varying the  density
ratios and the surface tensions independently and over a wide range of
parameters. The main limitation in achieving a systematic 
enhancement of density ratios is due to {\it spurious currents} 
in static curved interfaces. 
This limits both numerical  stability in the dynamical
evolution and the intimately physical correctness 
even for the static case. 

Second, we have shown how to overcome this problem by developing
improved versions of pseudo-potential interactions. The goal is to
reduce anisotropy contributions that are the source of spurious
currents. We achieved this systematically, either by a refinement of
the curved interface, so as to soften the local density gradients, or
by improving the isotropy of the discretized pseudo-potential. The
first method is more effective, leading to a numerical reduction of the maximal
current up to a factor $10$ with only a doubling in the grid
resolution.
We have shown that this 
stretching of the interface can be achieved by a simple rescaling of
the coupling strengths with the reference density of the
pseudopotential.  This permits to achieve an 'adaptive' form of local
grid refinement without changing the structure of the lattice nodes.

The present analysis has been carried out for a given choice of the
pseudopotential $\psi(\rho)$.  In principle, the major conclusions should
carry over to other, possibly more effective, functional forms of
$\psi$ \cite{Yuan}.

Besides clarifying the theoretical foundations of the original SC
model, it is hoped that the extended version presented in this work
will help setting the stage for future and more challenging
applications of pseudo-potential methods to the simulation of complex
multiphase flows.

\section{Appendix A}

In this appendix we discuss the tensorial structures that lead to a
vector structure of the form \be\label{FORCNOW}
J_{i}=\frac{1}{2}\psi \partial_{i} \Delta
\psi=\frac{1}{2}\psi \partial_{ill}\psi \ee where doubled indexes are
summed upon. We start from the most general expression for a second
order, non diagonal tensor involving derivatives only in the second
order: \be\label{general}
S_{ij}=\left[\frac{c}{2}(\partial_{l}\psi)(\partial_{l}\psi)+d\psi\partial_{ll}\psi
\right]\delta_{ij}+a(\partial_{i} \psi)(\partial_{j} \psi)+b
\psi \partial_{ij} \psi \ee where $a,b,c,d$ are meant to be fixed upon
consistency with expression (\ref{FORCNOW}). It is infact verified
that upon differentiation: \be
\partial_{j} S_{ij}=c\partial_{j} \psi \partial_{ij} \psi + d \partial_{i} \psi \partial_{jj} \psi +d \psi \partial_{ijj} \psi+ a \partial_{ij} \psi \partial_{j} \psi+a \partial_{i} \psi \partial_{jj} \psi + b \partial_{j} \psi \partial_{ij} \psi + b \psi \partial_{ijj}\psi.
\ee
To be consistent with the expression of the forcing we must impose:
\be\label{cond}
\left \{ \begin{array} {l} 
a+b+c=0   \\
a+d=0   \\
d+b=\frac{1}{2}   \\   \end{array} \right.
\ee
and we end up with three constraints and four constants. This means that there are infinitely many choices of $S_{ij}$ satisfying the condition:
\be
\partial_{j} S_{ij}= J_{i} \ee and we need another constraint to close
the problem and give unambiguously our tensor. Even if the tensor
structure is not uniquely determined, when we apply our arguments to
the case of a flat interface whose dishomogeneities develop along a $y$
coordinate, we notice that the normal component of the above tensor
is: 
\be\label{profile} S_{yy}=\frac{1}{2}\psi \partial_{yy}
\psi+\left( a+\frac{c}{2} \right) (\partial_{y} \psi)^2 \ee and from
the last $2$ expressions of (\ref{cond}) we obtain $a-b=-\frac{1}{2}$
that used in the first one imposes: \be 2a+c=-\frac{1}{2} \Rightarrow
a+\frac{c}{2}=-\frac{1}{4}.  \ee So, even if the tensor is not
uniquely determined, its normal component is uniquely 
given by 
\be \label{profile1} S_{yy}=\frac{1}{2}\psi \partial_{yy}
\psi-\frac{1}{4}(\partial_{y} \psi)^2.  \ee This implies that when
using a mechanical stability equation (\ref{nabla}) with a fixed
boundary condition we are able to extract the same profile as a
function of $y$. Then, from the expression (\ref{general}) we can also
write the equivalent of the surface tension considering the mismatch
between the normal $S_{yy}$ and tangential components $S_{xx}$ : \be
\sigma = \displaystyle\int_{-\infty}^{+\infty} ( a (\partial_{y}
\psi)^2+b \psi \partial_{yy}
\psi)=(a-b)\int_{-\infty}^{+\infty}(\partial_{y} \psi)^2.  \ee Again,
from that the last two expression of (\ref{cond}) we get
$a-b=-\frac{1}{2}$. This means that the surface tension is uniquely
determined.

\section{Appendix B}

In this appendix we show how to derive non isotropic contributions from discretizations. The forcing term is written in the form
\be
F_i=-{\cal G}_{b} c^{2}_{s} \psi({\bm x}) \sum_l w(|{\bm c}_l|^2)\psi({\bm x}+{\bm c}_l){c}_l^i.
\ee
Applying the Taylor expansion, one obtains
\begin{equation}
\begin{split}
F_{i}=-\G_{b} c^{2}_{s}\psi({\bm x})
\Biggl[E^{(2)}_{ij}\partial_{j}\psi
+\frac{1}{3!}E^{(4)}_{ijkl}\partial_{jkl}\psi
+\frac{1}{5!}E^{(6)}_{ijklpq}\partial_{jklpq}\psi
+\frac{1}{7!}E^{(8)}_{ijklpqst}\partial_{jklpqst}\psi
+
..
\Biggr],
\label{eq:f_tayl01a}
\end{split}
\end{equation}
with
\be
E^{(m)}=E^{(m)}_{i_{1}i_{2}...i_{m}}=\sum_{l}w(|{\bm c}_{l}|^2){c}^{i_{1}}_{l}{c}^{i_2}_{l}...{c}^{i_m}_{l}.
\ee
and (obviously) zero odd tensors:
\be
E^{(2n+1)}=0.
\ee
The even tensors are written as
\be
E_{i_1i_2...i_{2n}}^{(2n)}
={\cal C}^{(2n)}\Delta_{i_1i_2...i_{2n}}^{(2n)},
\label{eq:iso_2n}
\ee
where $\Delta^{(2n)}$ is given by the recursion relation \cite{Wolfram}
$$
\Delta_{ij}^{(2)}=\delta_{ij},
\ \ \ 
\Delta_{ijkl}^{(4)}=
\delta_{ij}\delta_{kl}+
\delta_{ik}\delta_{jl}+
\delta_{il}\delta_{jk},
$$
\be \Delta_{i_1i_2...i_{2n}}^{(2n)} = \sum_{j=2}^{2n} \delta_{i_1 i_j}
\Delta_{i_2 i_3...i_{j-1}i_{j+1}...i_{2n}}^{(2n-2)}.  \ee In our mean
field approach, $\psi$ is a function of the density.  If the density
distribution is axially symmetric, $\psi$ is also axially symmetric,
$\psi=\psi(r)$.  Then, the force $F_i$ should be written as \be
F_i({\bm x})=\frac{x_i}{r}\tilde{F}(r),
\label{eq:fi_iso01}
\ee
where $\tilde{F}$ is a scalar function.
It should be noted that
the isotropy (\ref{eq:iso_2n}) for all ${\bm E}^{(2n)}$ is essential
in order to satisfy the relation (\ref{eq:fi_iso01}). Now, we will show 
that the truncated isotropy induces the anisotropic force, which triggers the spurious currents,  even when the density distribution is axially symmetric.
Let us consider the standard case of $2DQ9$.  As already noticed in the text, this is a  $4$th-order approximation in the isotropy and the weights are given by
$$
w(1)=1/3,\ \ w(2)=1/12,
$$
\be
w(n)=0\ \ \ {\rm for}\ \ \ n\ge 3.
\ee
This approximation means that 
all the tensors up to the $4$th-order 
(${\bm E}^{(2)}$ and ${\bm E}^{(4)}$) 
are isotropic but the higher order ones
${\bm E}^{(n)}$ ($n\geq 6$) are not.
Using standard Taylor expansion for lattice Boltzmann populations one
obtains after some lengthly algebra:
\begin{equation}
\begin{split}
\sum_l w(|{\bm c}_l|^2)\psi({\bm x}+{\bm c}_l)c_l^x
=&\partial_x \psi+\frac{1}{6}\partial_x(\partial_x^2+\partial_y^2)\psi
+\partial_x\left(
\frac{1}{120}\partial_x^4+\frac{1}{36}\partial_x^2\partial_y^2
+\frac{1}{72}\partial_y^4
\right)\psi+...,
\\
\sum_l w(|{\bm c}_l|^2)\psi({\bm x}+{\bm c}_l)c_l^y
=&\partial_y \psi+\frac{1}{6}\partial_y(\partial_x^2+\partial_y^2)\psi
+\partial_y\left(
\frac{1}{72}\partial_x^4+\frac{1}{36}\partial_x^2\partial_y^2
+\frac{1}{120}\partial_y^4
\right)\psi+...
\label{eq:wpsic01}
\end{split}
\end{equation}
Using a nabla operator, (\ref{eq:wpsic01}) is rewritten as
\be
\sum_l w(|{\bm c}_l|^2)\psi({\bm x}+{\bm c}_l){\bm c}_l
=
{\bm \nabla}
\left(
1+\frac{1}{6}\nabla^2
+\frac{1}{72}\nabla^2\nabla^2
\right)\psi
+{\bm e}_x\frac{\partial_x^5\psi}{180}
+{\bm e}_y\frac{\partial_y^5\psi}{180}
+
{\cal O}(\partial^7),
\label{eq:wpsic02}
\ee
where ${\bm e}_x=(1,0)$ and ${\bm e}_y=(0,1)$ are unit vectors in Cartesian coordinates. 
Next we assume axial symmetry of the density distribution, 
i.e., $\psi=\psi(r)$. Because of 
${\bm \nabla}={\bm e}_r\partial r$ and 
$\nabla^2=\partial_r^2+r^{-1}\partial_r$
for an axially symmetric function, 
the 1st term in the r.h.s of (\ref{eq:wpsic02})
is isotropic. 
On the other hand, 
the 2nd and 3rd terms, 
related to the anisotropic tensor ${\bm E}^{(6)}$ in (\ref{eq:f_tayl01a}),
are not. 
Now the force $F_i$ is decomposed into 
the isotropic and anisotropic parts, i.e.,
\be
F_i({\bm x})=\frac{x_i}{r}\tilde{F}(r)+F_i'({\bm x}).
\ee
Within the $O(\partial^5)$ approximation, 
$\tilde{F}(r)$ and $F_i'({\bm x})$
are respectively given by 
\be
\tilde{F}(r)
=
-{\cal G}_{b}c_{s}^{2}\psi
\partial_r\Biggl(
1+\frac{1}{6r}
\frac{{\rm d}}{{\rm d}r}r\frac{{\rm d}}{{\rm d}r}
+\frac{1}{72r}
\frac{{\rm d}}{{\rm d}r}r\frac{{\rm d}}{{\rm d}r}
r^{-1} \frac{{\rm d}}{{\rm d}r}r\frac{{\rm d}}{{\rm d}r}
\Biggr)\psi,
\ee
\be
F_1'=-\frac{{\cal G}_{b} c_{s}^{2}\psi\partial_x^5\psi}{180},
\ \ \ 
F_2'=-\frac{{\cal G}_{b} c_{s}^{2}\psi\partial_y^5\psi}{180}.
\label{eq:aniso_force01}
\ee The anisotropic force $F_i'$ due to the anisotropy of ${\bm
  E}^{(6)}$ is responsible for the spurious currents. Higher orders
can be computed similarly (for the interested reader please contact
the authors).

\section{Appendix C}

Here we detail the exact procedures leading to higher order isotropic
terms in the forcing contribution for a regular lattice in $2d$. The
velocity phase space and forcing weights for isotropic terms  up to
$16$th order are explicitly given in figure (\ref{fig:6}). To treat 
correctly isotropy from a lattice set of velocity vectors
(${c}^{i}_{l}, i=x,y$) the starting point is the $2$ point tensor
on the lattice which is assumed normalized to unity 
\be
\sum_{l}{c}_l^i{c}_l^j w(|{\bm c}_{l}|^2)=\delta_{ij}.
\label{eq:iso2nd}
\ee
Considering the regular structure of the lattice and the consequent symmetry of $c_l^i$ with respect to $i=x$ and $i=y$,
one can write (\ref{eq:iso2nd}) in the simplified form
\be
\sum_{l} (c_l^x)^2 w(|{\bm c}_{l}|^2)=1.
\label{eq:iso2nd2}
\ee
The fourth-order isotropy is imposed by 
\be
\sum_{l}{c}_l^i{c}_l^j{c}_l^j{c}_l^s w(|{\bm c}_{l}|^2)=
{\cal C}^{(4)}\left(
\delta_{ij}\delta_{ks}
+
\delta_{ik}\delta_{js}
+
\delta_{is}\delta_{jk}
\right),
\label{eq:iso4th}
\ee
where ${\cal C}^{(4)}$ is a constant.
Since one obtains
$$
\sum_l ({c}_l^x)^4 w(|{\bm c}|^2_{l})=3{\cal C}^{(4)},
\ \ \ 
\sum_l ({c}_l^x)^2({c}_l^y)^2 w(|{\bm c}_{l}|^2)={\cal C}^{(4)},
\label{eq:iso4th2}
$$
a condition to satisfy (\ref{eq:iso4th}) is written as
\be
\frac{\sum_l ({c}_l^x)^4 w(|{\bm c}_{l}|^2)}
{\sum_l ({c}_l^x)^2({c}_l^y)^2 w(|{\bm c}_{l}|^2)}=3.
\label{eq:iso4th3}
\ee
In terms of lattice vector this can be achieved with the standard $2DQ9$ model with weights $w(1)=1/3$ and $w(2)=1/12$ and the corresponding lattice velocities:\\
\noindent
$w(1)$:
\be
({\bm c}_1\ 
{\bm c}_2\ 
{\bm c}_3\ 
{\bm c}_4)
=
\left(
\begin{array}{rrrr}
1&0&-1&0\\
0&1&0&-1\\
\end{array}
\right),
\ee
\noindent
$w(2)$:
\be
(
{\bm c}_5\ 
{\bm c}_6\ 
{\bm c}_7\ 
{\bm c}_8)
=
\left(
\begin{array}{rrrr}
1&-1&-1& 1\\
1& 1&-1&-1\\
\end{array}
\right).
\ee
\noindent
More general conditions can then be obtained for higher order tensors. For example, the $6$th, $8$th, $10$th and $12$th-order isotropies are given by
\be
\begin{split}
\sum_{l}{c}_l^{i_1}{c}_l^{i_2}{c}_l^{i_3}{c}_l^{i_4}{c}_l^{i_5}{c}_l^{i_6}w(|{\bm c}_{l}|^2)=&
{\cal C}^{(6)}\left(
\delta_{i_1i_2}\delta_{i_3i_4}\delta_{i_5i_6}
+
..
\right),
\\
\sum_{l}{c}_l^{i_1}{c}_l^{i_2}{c}_l^{i_3}{c}_l^{i_4}{c}_l^{i_5}{c}_l^{i_6}{c}_l^{i_7}{c}_l^{i_8}w(|{\bm c}_{l}|^2)=&
{\cal C}^{(8)}\left(
\delta_{i_1i_2}\delta_{i_3i_4}\delta_{i_5i_6}\delta_{i_7i_8}
+
..
\right),
\\
\sum_{l}{c}_l^{i_1}{c}_l^{i_2}{c}_l^{i_3}{c}_l^{i_4}{c}_l^{i_5}{c}_l^{i_6}{c}_l^{i_7}{c}_l^{i_8}{c}_l^{i_9}{c}_l^{i_{10}}w(|{\bm c}_{l}|^2)=&
{\cal C}^{(10)}\left(
\delta_{i_1i_2}\delta_{i_3i_4}\delta_{i_5i_6}\delta_{i_7i_8}\delta_{i_9i_{10}}
+
..
\right),
\\
\sum_{l}{c}_l^{i_1}{c}_l^{i_2}{c}_l^{i_3}{c}_l^{i_4}{c}_l^{i_5}{c}_l^{i_6}{c}_l^{i_7}{c}_l^{i_8}
{c}_l^{i_9}{c}_l^{i_{10}}{c}_l^{i_{11}}{c}_l^{i_{12}}w(|{\bm c}_{l}|^2)=&
{\cal C}^{(12)}\left(
\delta_{i_1i_2}\delta_{i_3i_4}\delta_{i_5i_6}\delta_{i_7i_8}\delta_{i_9i_{10}}\delta_{i_{11}i_{12}}
+
..
\right).
\end{split}
\ee And the mixed contributions can be constructed as well: 
\be
\sum_l ({c}_l^x)^{2n}({c}_l^y)^{2m} w(|{\bm c}_{l}|^2)={\cal
  C}^{(2n+2m)}(2n-1)!!(2m-1)!!, \ee where
$(2n-1)!!=(2n-1)\times(2n-3)\times ...\times 1$.  Then, to achieve
isotropy at higher orders one should introduce some requirements on
the tensors. Just to give an example, for the isotropy up to the $6$th
order one should require that: 
\be 
\frac{\sum_l({c}_l^x)^4 w(|{\bm c}_{l}|^2) }
{\sum_l ({c}_l^x)^2({c}_l^y)^2 w(|{\bm c}_{l}|^2)}=3
\label{eq:iso4th3a}
\ee
\be
\frac{\sum_l ({c}_l^x)^6 w(|{\bm c}_{l}|^2)}{\sum_l ({c}_l^x)^4({c}_l^y)^2 w(|{\bm c}_{l}|^2)}=5.
\label{eq:iso_6_1}
\ee
these translate to  the matrix relation
\be
\left(
\begin{array}{rrr}
2&  4&   8\\
1& -4&  16\\
1& -8&  64\\
\end{array}
\right)
\left(
\begin{array}{l}
 w(1)\\
 w(2)\\
 w(4)\\
\end{array}
\right)
=
\left(
\begin{array}{l}
1\\
0\\
0\\
\end{array}
\right)
\ee
that can be satisfied using $12$ velocities 
with weights $w(1)=4/15,w(2)=1/10$ and $w(4)=1/120$\\

\noindent
$w(1)$:
\be
({\bm c}_1\ 
{\bm c}_2\ 
{\bm c}_3\ 
{\bm c}_4)
=
\left(
\begin{array}{rrrr}
1&0&-1&0\\
0&1&0&-1\\
\end{array}
\right),
\ee
\noindent
$w(2)$:
\be
(
{\bm c}_5\ 
{\bm c}_6\ 
{\bm c}_7\ 
{\bm c}_8)
=
\left(
\begin{array}{rrrr}
1&-1&-1& 1\\
1& 1&-1&-1\\
\end{array}
\right),
\ee
\noindent
$w(4)$:
\be
(
{\bm c}_{9}\ 
{\bm c}_{10}\ 
{\bm c}_{11}\ 
{\bm c}_{12})
=
\left(
\begin{array}{rrrr}
2&0&-2& 0\\
0&2& 0&-2\\
\end{array}
\right).
\ee
\noindent
Higher order calculations are lengthly and not reported here. The set of
vectors can be extracted from figure (\ref{fig:6}) while the 
weights can be found in table 1.

\section{Appendix D}
\maketitle

The same calculations are then arranged in $3d$. For each $w(n)$ (reported in table 2), 
the corresponding velocity vectors ${\bm c}_l$ are shown below:

\vspace{1em}

\noindent
$w(1)        $:
\be
(
{\bm c}_{1}\ 
{\bm c}_{2}\ 
{\bm c}_{3}\ 
{\bm c}_{4}\
{\bm c}_{5}\ 
{\bm c}_{6})
=
\left(
\begin{array}{rrrrrrrr}
1&-1&0& 0&0& 0\\
0& 0&1&-1&0& 0\\
0& 0&0& 0&1&-1\\
\end{array}
\right),
\label{eq:cw1}
\ee
\noindent
$w(2)        $:
\be
\begin{split}
&(
{\bm c}_{ 7}\ 
{\bm c}_{ 8}\ 
{\bm c}_{ 9}\ 
{\bm c}_{10}\
{\bm c}_{11}\ 
{\bm c}_{12}\
{\bm c}_{13}\ 
{\bm c}_{14}\ 
{\bm c}_{15}\ 
{\bm c}_{16}\
{\bm c}_{17}\ 
{\bm c}_{18})
\\=&
\left(
\begin{array}{rrrrrrrrrrrrrrrr}
1& 1&-1&-1&1& 1&-1&-1&0& 0& 0& 0\\
1&-1& 1&-1&0& 0& 0& 0&1& 1&-1&-1\\
0& 0& 0& 0&1&-1& 1&-1&1&-1& 1&-1\\
\end{array}
\right),
\end{split}
\ee
\noindent
$w(3)        $:
\be
(
{\bm c}_{19}\ 
{\bm c}_{10}\ 
{\bm c}_{21}\ 
{\bm c}_{22}\
{\bm c}_{23}\ 
{\bm c}_{24}\
{\bm c}_{25}\ 
{\bm c}_{26})
=
\left(
\begin{array}{rrrrrrrrrr}
1& 1& 1& 1&-1&-1&-1&-1\\
1& 1&-1&-1& 1& 1&-1&-1\\
1&-1& 1&-1& 1& 1&-1&-1\\
\end{array}
\right),
\ee
\noindent
$w(4)        $:
\be
(
{\bm c}_{27}\ 
{\bm c}_{28}\ 
{\bm c}_{29}\ 
{\bm c}_{30}\
{\bm c}_{31}\ 
{\bm c}_{32})
=
\left(
\begin{array}{rrrrrrrr}
2&-2&0& 0&0& 0\\
0& 0&2&-2&0& 0\\
0& 0&0& 0&2&-2\\
\end{array}
\right),
\ee
\noindent
$w(5)        $:
\be
\begin{split}
&(
{\bm c}_{33}\ 
{\bm c}_{34}\ 
{\bm c}_{35}\ 
{\bm c}_{36}\
{\bm c}_{37}\ 
{\bm c}_{38}\
{\bm c}_{39}\ 
{\bm c}_{40}\ 
{\bm c}_{41}\ 
{\bm c}_{42}\
{\bm c}_{43}\ 
{\bm c}_{44}\\& 
{\bm c}_{45}\ 
{\bm c}_{46}\ 
{\bm c}_{47}\ 
{\bm c}_{48}\ 
{\bm c}_{49}\ 
{\bm c}_{50}\
{\bm c}_{51}\ 
{\bm c}_{52}\ 
{\bm c}_{53}\ 
{\bm c}_{54}\
{\bm c}_{55}\ 
{\bm c}_{56})
\\=&
\left(
\begin{array}{rrrrrrrrrrrr}
2& 2&-2&-2&2& 2&-2&-2&1&-1& 1&-1\\
1&-1& 1&-1&0& 0& 0& 0&2& 2&-2&-2\\
0& 0& 0& 0&1&-1& 1&-1&0& 0& 0& 0\\
\end{array}
\right.\\&\left.
\begin{array}{rrrrrrrrrrrr}
0& 0& 0& 0&1&-1& 1&-1&0& 0& 0& 0\\
2& 2&-2&-2&0& 0& 0& 0&1&-1& 1&-1\\
1&-1& 1&-1&2& 2&-2&-2&2& 2&-2&-2\\
\end{array}
\right),
\end{split}
\ee
\noindent
$w(6)        $:
\be
(
{\bm c}_{57}\ 
..\ 
{\bm c}_{80})
=
\left(
\begin{array}{rrr}
2&...&-1\\
1&...&-1\\
1&...&-2\\
\end{array}
\right),
\ee
\noindent
$w(8)        $:
\be
(
{\bm c}_{81}\ 
..\ 
{\bm c}_{92})
=
\left(
\begin{array}{rrr}
2&...& 0\\
2&...&-2\\
0&...&-2\\
\end{array}
\right),
\ee
\noindent

\noindent
$w_{221}(9)  $:
\be
(
{\bm c}_{93}\ 
..\ 
{\bm c}_{116})
=
\left(
\begin{array}{rrr}
2&...&-1\\
2&...&-2\\
1&...&-2\\
\end{array}
\right),
\ee
\noindent
$w_{300}(9)  $:
\be
(
{\bm c}_{117}\ 
..\ 
{\bm c}_{122})
=
\left(
\begin{array}{rrr}
3&...& 0\\
0&...& 0\\
0&...&-3\\
\end{array}
\right),
\ee
\noindent
$w(10)       $:
\be
(
{\bm c}_{123}\ 
..\ 
{\bm c}_{146})
=
\left(
\begin{array}{rrr}
3&...& 0\\
1&...&-1\\
0&...&-3\\
\end{array}
\right),
\ee
\noindent
$w(11)       $:
\be
(
{\bm c}_{147}\ 
..\ 
{\bm c}_{170})
=
\left(
\begin{array}{rrr}
3&...&-1\\
1&...&-1\\
1&...&-3\\
\end{array}
\right).
\ee

\newpage


\newpage

\newpage
\newpage
\begin{figure}[t]
\begin{center}
\includegraphics[scale=0.5]{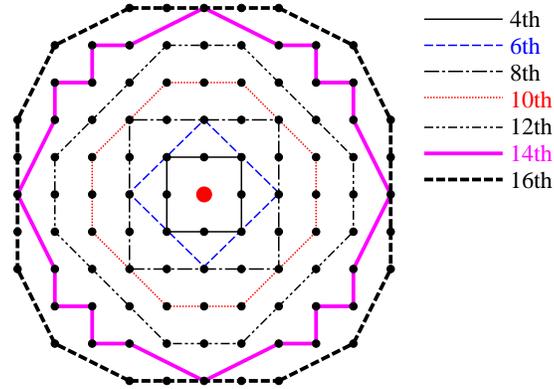}
\caption{ The grid points identifying the set of velocity fields for a
  $2d$ case. With reference to the weights reported in table 1,
  different degrees of isotropy can be achieved: 4th order (up to
  $w(2)$), 6th order (up to $w(4)$), 8th order (up to $w(8)$), 10th
  order (up to $w(10)$), 12th order (up to $w(17)$), 14th order (up to
  $w(25)$) and 16th order (up to $w(32)$).  }
\label{fig:6}
\end{center}
\end{figure}

\begin{figure}
  \begin{center}
    \includegraphics[scale=.7]{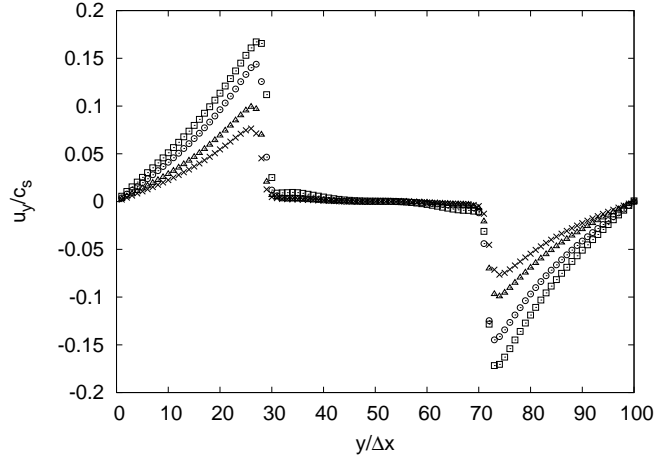}
    \caption{Reduction of spurious currents with higher order
      isotropic tensors in the forcing terms. The spurious
      contributions for a static drop are analyzed for a density ratio
      $\Delta \rho/\rho_{g} \sim 50$ obtained for a standard SC model
      (\ref{forcing}) and (\ref{PSI}). The parameter chosen is ${\cal
        G}_{b}=-7.0$. We show the vertical velocity, $u_{y}$,
      normalized with the lattice speed of sound, $c_{s}$, at
      $x=L_{x}/2$ and as a function of $y/\Delta x$. The different
      plots correspond to different degrees of isotropy in the forcing
      term (\ref{forcing}): $4$th order ($\Box$), $6$th order
      ($\circ$), $8$th order ($\triangle$) and $10$th order
      ($\times$). Notice the reduction of the spurious contributions in
      the proximity of the droplet surface, $y/\Delta x \sim
      30,70$. Data are obtained from Lattice Boltzmann equation
      (\ref{eq:LB}) in a fully periodic domain $L_{x} \times L_{y} =
      100 \Delta x \times 100 \Delta x$ and $\tau=1.0 \Delta t$ in
      lattice units.}
\label{fig:7}
  \end{center}
\end{figure}

\newpage
\begin{figure}
  \begin{center}
    \includegraphics[scale=.7]{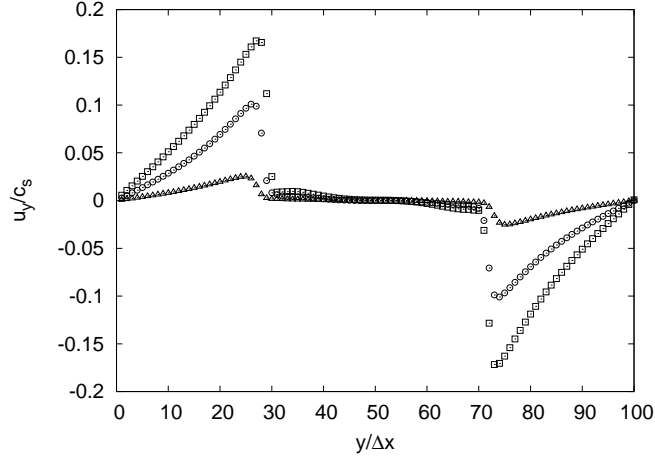}
    \caption{reduction of the spurious currents with grid refinement
      at changing $\rho_0$. The spurious contributions for a static
      drop are analyzed for a density ratio $\Delta \rho/\rho_{g} \sim
      50$ obtained for the improved SC model (\ref{forcing2}) and
      (\ref{PSINEW}). The parameters are chosen in such a way to
      reproduce (\ref{TENSORENEW}) with $A_{1}=-7.0$. We show the
      vertical velocity, $u_{y}$, normalized with the lattice speed of
      sound, $c_{s}$, at $x=L_{x}/2$ ad as a function of $y/\Delta
      x$. The different plots correspond to different degrees of
      refinement obtained with $A_{2}=-7.0/\rho^2_{0}$ in
      (\ref{TENSORENEW}): $\rho_{0}=1.0$ ($\Box$), $\rho_{0}=0.75$
      ($\circ$) and $\rho_{0}=0.5$ ($\triangle$). Details of the
      numerical simulations are the same of the previous figure.}
 \label{fig:8}
  \end{center}
\end{figure}
\begin{figure}
  \begin{center}
    \includegraphics[scale=.7]{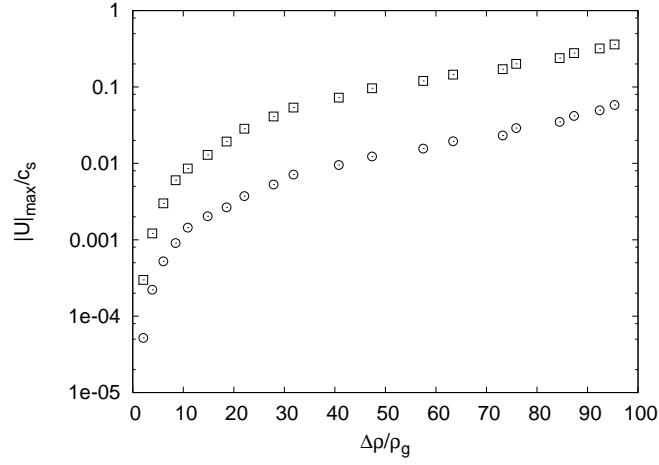}
    \caption{Reduction of the spurious currents with refinement in the
      forcing terms. The spurious contributions for a static drop are
      analyzed for various density ratios obtained in the improved SC
      model (\ref{forcing2}) and (\ref{PSINEW}). The parameters are
      chosen in such a way as to reproduce (\ref{TENSORENEW}) with
      different $A_{1}$, thus changing the density ratio $\Delta
      \rho/\rho_{g}$. We show the maximum velocity due to spurious
      contributions, normalized with the lattice speed of sound
      ($c_{s}$) as a function of $\Delta \rho/\rho_{g}$. The $2$
      different plots correspond to different degree of refinement
      obtained with $A_{2}=7.0/\rho^2_{0}$ in (\ref{eq:rescaled}) and
      different values of $\rho_{0}$: $\rho_{0}=1.0$ ($\Box$),
      $\rho_{0}=0.5$ ($\circ$). Notice the net reduction of the
      spurious contributions at fixed macroscopic physics (same
      density ratio and surface tension) obtained through the
      rescaling $A_{2} \sim 1/ \rho_{0}^2$. The numerical details are
      the same described in figure (\ref{fig:8}).}
    \label{fig:9}
  \end{center}
\end{figure}


\clearpage
\newpage

\begin{center}

\def\arraystretch{2.2}

Table 1. Weights up to the 16th-order approximation for the case of
$2d$ models.
Notice that the weights for velocities with $|{\bm c}_l|^2=25$ needs to be
chosen differently according to the directions in the $x-y$ plane. The
notation   $w_{x, y}(|{\bm c}_l|^2)$ stands for the
velocity lattice vectors $(\pm x_1, \pm x_2 )$ and  $(\pm x_2, \pm x_1)$.

\begin{tabular}{|c|cccccc|}
\hline
 &${\bm E}^{(6)}$&${\bm E}^{(8)}$&${\bm E}^{(10)}$&${\bm E}^{(12)}$&${\bm E}^{(14)}$&${\bm E}^{(16)}$\\\hline
 $w(1)$     &$\displaystyle\frac{4}{ 15}$&$\displaystyle\frac{4}{  21}$&$\displaystyle\frac{262}{1785}$&$\displaystyle\frac{68}{   585}$&$\displaystyle\frac{ 19414}{  228375}$&$\displaystyle\frac{ 285860656}{  3979934595}$\\
 $w(2)$     &$\displaystyle\frac{1}{ 10}$&$\displaystyle\frac{4}{  45}$&$\displaystyle\frac{ 93}{1190}$&$\displaystyle\frac{68}{  1001}$&$\displaystyle\frac{549797}{10048500}$&$\displaystyle\frac{2113732952}{ 43779280545}$\\
 $w(4)$     &$\displaystyle\frac{1}{120}$&$\displaystyle\frac{1}{  60}$&$\displaystyle\frac{  7}{ 340}$&$\displaystyle\frac{ 1}{    45}$&$\displaystyle\frac{175729}{ 7917000}$&$\displaystyle\frac{ 940787801}{ 43779280545}$\\
 $w(5)$     &                            &$\displaystyle\frac{2}{ 315}$&$\displaystyle\frac{  6}{ 595}$&$\displaystyle\frac{62}{  5005}$&$\displaystyle\frac{ 50728}{ 3628625}$&$\displaystyle\frac{ 124525000}{  8755856109}$\\
 $w(8)$     &                            &$\displaystyle\frac{1}{5040}$&$\displaystyle\frac{  9}{9520}$&$\displaystyle\frac{ 1}{   520}$&$\displaystyle\frac{  3029}{  913500}$&$\displaystyle\frac{  15841927}{  3979934595}$\\
 $w(9)$     &                            &                             &$\displaystyle\frac{  2}{5355}$&$\displaystyle\frac{ 4}{  4095}$&$\displaystyle\frac{ 15181}{ 7536375}$&$\displaystyle\frac{   2046152}{   795986919}$\\
$w(10)$     &                            &                             &$\displaystyle\frac{  1}{7140}$&$\displaystyle\frac{ 2}{  4095}$&$\displaystyle\frac{   221}{  182700}$&$\displaystyle\frac{  14436304}{  8755856109}$\\
$w(13)$     &                            &                             &                               &$\displaystyle\frac{ 2}{ 45045}$&$\displaystyle\frac{    68}{  279125}$&$\displaystyle\frac{  18185828}{ 43779280545}$\\
$w(16)$     &                            &                             &                               &$\displaystyle\frac{ 1}{480480}$&$\displaystyle\frac{  1139}{26796000}$&$\displaystyle\frac{  13537939}{140093697744}$\\
$w(17)$     &                            &                             &                               &                             $0$&$\displaystyle\frac{    68}{ 2968875}$&$\displaystyle\frac{    231568}{  3979934595}$\\
$w(18)$     &                            &                             &                               &                                &$\displaystyle\frac{    17}{ 1425060}$&$\displaystyle\frac{   1516472}{ 43779280545}$\\
$w(20)$     &                            &                             &                               &                                &$\displaystyle\frac{    17}{ 5742000}$&$\displaystyle\frac{     18769}{  1591973838}$\\
$w_{50}(25)$&                            &                             &                               &                                &$\displaystyle\frac{     1}{32657625}$&$\displaystyle\frac{       184}{   315867825}$\\
$w_{34}(25)$&                            &                             &                               &                                &$\displaystyle\frac{     1}{32657625}$&$\displaystyle\frac{       464}{   795986919}$\\
$w(26)$     &                            &                             &                               &                                &                                      &$\displaystyle\frac{      1448}{  4864364505}$\\
$w(29)$     &                            &                             &                               &                                &                                      &$\displaystyle\frac{       148}{  4864364505}$\\ 
$w(32)$     &                            &                             &                               &                                &                                      &$\displaystyle\frac{       629}{400267707840}$\\ \hline 
\end{tabular}
\end{center}

\newpage

\begin{center}

\def\arraystretch{2.2}

Table 2\ \ \ 
Weights up to the 10th-order approximation for $3d$ models. Notice that the weights for velocities with $|{\bm c}_l|^2=9$ needs to be
chosen differently according to the directions in the $x-y-z$ space. The
notation   $w_{x, y, z}(|{\bm c}_l|^2)$ stands for the 
velocity lattice vectors $(\pm x_1, \pm x_2, \pm x_3 )$ plus
permutation.

\begin{tabular}{|c|ccc|}
\hline
              &$\displaystyle{\bm E}^{(6)}$&$\displaystyle{\bm E}^{(8)} $&$\displaystyle{\bm E}^{(10)}   $\\\hline
$w(1)        $&$\displaystyle\frac{2}{ 15}$&$\displaystyle\frac{4}{  45}$&$\displaystyle\frac{352}{ 5355}$\\
$w(2)        $&$\displaystyle\frac{1}{ 15}$&$\displaystyle\frac{1}{  21}$&$\displaystyle\frac{ 38}{ 1071}$\\
$w(3)        $&$\displaystyle\frac{1}{ 60}$&$\displaystyle\frac{2}{ 105}$&$\displaystyle\frac{271}{14280}$\\
$w(4)        $&$\displaystyle\frac{1}{120}$&$\displaystyle\frac{5}{ 504}$&$\displaystyle\frac{139}{14280}$\\
$w(5)        $&$\displaystyle             $&$\displaystyle\frac{1}{ 315}$&$\displaystyle\frac{ 53}{10710}$\\
$w(6)        $&$\displaystyle             $&$\displaystyle\frac{1}{ 630}$&$\displaystyle\frac{  5}{ 2142}$\\
$w(8)        $&$\displaystyle             $&$\displaystyle\frac{1}{5040}$&$\displaystyle\frac{ 41}{85680}$\\
$w_{221}(9)  $&$\displaystyle             $&$\displaystyle              $&$\displaystyle\frac{  1}{ 4284}$\\
$w_{300}(9)  $&$\displaystyle             $&$\displaystyle              $&$\displaystyle\frac{  1}{ 5355}$\\
$w(10)       $&$\displaystyle             $&$\displaystyle              $&$\displaystyle\frac{  1}{10710}$\\
$w(11)       $&$\displaystyle             $&$\displaystyle              $&$\displaystyle\frac{  1}{42840}$\\
\hline
\end{tabular}
\end{center}

\end{document}